\documentclass{article}

\usepackage{amsmath}
\usepackage{graphicx}
\usepackage{cite}

\title{Thin film dynamics with surfactant phase transition}

\author{M. H. K\"opf\thanks{E-mail: m.koepf@uni-muenster.de}
  \and S. V. Gurevich \and R. Friedrich}
\date{}
\begin{document}
\maketitle
\begin{center}
{\it Institute for Theoretical Physics, University of M\"unster -
    Wilhelm-Klemm-Stra{\ss}e 9, D-48149 M\"unster, Germany}
\end{center}
\begin{abstract}
A thin liquid film covered with an insoluble surfactant in the
vicinity of a first-order phase transition is discussed. Within the
lubrication approximation we derive two coupled equations to
describe the height profile of the film and the surfactant density.
Thermodynamics of the surfactant is incorporated via a
Cahn-Hilliard type free-energy functional which can be chosen to
describe a transition between two stable phases of different
surfactant density. Within this model, a linear stability analysis
of stationary homogeneous solutions is performed, and 
drop formation in a film covered with surfactant in the lower
density phase is investigated numerically in one and two spatial
dimensions.
\end{abstract}
\section{Introduction}
The stability and dynamics of thin liquid films have been of
considerable interest to both experimental and theoretical
research \cite{ODB_RevModPhys_97,dG_RevModPhys_85,PP_PRE_00,R_PRL_92}. 
When the thickness of a flat liquid film is in the range of
$\sim100\,\mathrm{nm}$, it becomes sensitive to interaction with its
substrate. In a certain range of film thickness determined by the exact
form of the interaction potential, this will render the film unstable with
respect to small perturbations and a pattern formation process sets in.
Depending on its initial height, the film breaks up into droplets,
labyrinth-like patterns or arrays of holes
\cite{BN_PRL_01,R_PRL_92}. This process is known as spinodal
dewetting.

Brought onto the surface of a liquid film, an insoluble surfactant, for
example an organic molecule with a hydrophilic head group and a
hydrophobic tail group, alters the surface tension and thereby
influences the breakup process. In addition, gradients of surfactant
density lead to so-called Marangoni convection on the surface resulting
in new instabilities like surfactant-induced fingering
\cite{TWS_PRL_89,CM_PhysFluids_06}. 

Most surfactants exhibit complicated thermodynamics with several phase
transitions \cite{KMD_RevModPhys_99,AGS_JPhysique_78}. These affect thin film
hydrodynamics via an equation of state, relating surface tension to
surfactant density \cite{RL_JPhysChemB_98}. In studies of the dynamics
of surfactant covered thin films, surfactant thermodynamics has so far
not been paid much attention to, because the hydrodynamics is
dominated by Marangoni convection rather than by effects of lateral
pressure and diffusion \cite{MT_PhysFluids_97}.

However, there is a special focus of experimental research on pattern
formation under conditions close to the so-called main transition in
monolayers of lipids like pulmonary surfactant
Dipalmitoylphosphatidylcholine (DPPC)
\cite{GCF_Nature_00,KCR_EPS_94,CLH_AccChemRes_07}. In vicinity of this
first-order phase transition parts of the surfactant in the
liquid-expanded (LE) phase and in the liquid-condensed (LC) phase
coexist. In these experiments, a substrate is coated with a lipid
monolayer via Langmuir-Blodgett transfer, i.e. it is withdrawn from a
trough filled with water on which a lipid monolayer has been prepared.
The observed patterns consist of ordered arrays of LE and LC domains,
including regular stripes and rectangles, the formation of
which is usually attributed to oscillations of the meniscus between the
water in the trough and the substrate \cite{CLH_AccChemRes_07}. A full
understanding of these phenomena cannot be achieved without
understanding the dynamics of the film and the surfactant near the main
transition. The aim of this letter is to outline theoretical
description of the evolution of a thin film covered with a surfactant
undergoing a phase transition.

The model we are going to present is derived within the lubrication
approximation \cite{ODB_RevModPhys_97}. We follow the usual approach
\cite{GG_JFluidMech_90,CM_PhysFluids_06,WCM_PhysFluids_02,DP_JCollIntSci_84,WGC_PhysFluids_94}
to describe the time evolution of the surfactant covered thin film by
two coupled partial differential equations, describing the height profile of
the underlying liquid film and the surfactant density.  The
surfactant phase transition is incorporated by choice of a suitable
free-energy functional which determines the lateral pressure as well as the
diffusive flux.
We are going to perform a linear stability analysis of stationary
homogeneous solutions of the derived equations and investigate the
effect of the surfactant on drop formation by numerical simulations. 
\begin{figure}
\begin{center}
\includegraphics{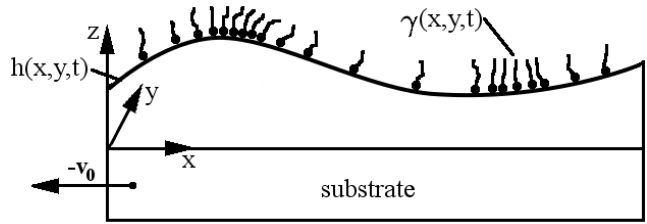}
\caption{Schematic of a surfactant-laden thin film of water on a
  substrate. Height profile $h(x,y,t)$ indicates the film thickness at
    location $(x,y)$ and time $t$, whereas $\gamma(x,y,t)$ describes the
    surfactant density at the surface above $(x,y)$.}
\label{fig.profile}
\end{center}
\end{figure}
\section{Lubrication approximation}
We consider a thin liquid film covered with an insoluble surfactant on a
moving solid substrate (see fig.~\ref{fig.profile}). 
{The velocity field of the liquid film can be obtained within the
lubrication approximation \cite{ODB_RevModPhys_97}. By this procedure
the initially three-dimensional flow problem is reduced to an effectively
two-dimensional one. The liquid film is then described by a height
profile $h(x,y,t)$, which indicates the local film thickness, and the
two-dimensional flow field at the surface ${\bf \tilde{u}}(x,y,t)$.
The surfactant density at the surface above the point $(x,y)$ is
described by the function $\gamma(x,y,t)$. The continuity equation of an
insoluble surfactant has been the subject of considerable discussion
\cite{S_ChemEngSci_60, S_PhysFluidsA_90, Aris}. Surface geometry is of
negligible influence in the lubrication regime, leaving us with the
nondimensionalized conservation law
\begin{equation}
\Gamma_{,T} = -\nabla\cdot\left[\Gamma{\bf \tilde{U}}+{\bf I}
\right]\,, \label{e.evolgamma}
\end{equation}
where ${\bf I}$ is the diffusive flux of the surfactant.
Here we have scaled all quantities by characteristic values:
\begin{align}
& X=\frac{\epsilon x}{h_0}\,, \qquad Y=\frac{\epsilon y}{h_0}\,, \qquad
T=\frac{\epsilon u_0 t}{h_0}\,, \notag \\ 
& {\bf U}=\frac{{\bf u}}{u_0}\,, \qquad
\Gamma=\frac{\gamma}{\gamma_0}\,, \qquad {\bf I}=
\frac{{\bf i}}{\gamma_0 u_0}\,,
\end{align}
and $\nabla:=(\partial_X,\partial_Y)$ denotes the nabla operator in
non-dimensional coordinates $X,Y$.  The dimensionless parameter
$\epsilon=h_0/l_0$ defines the ratio of characteristic height and length
scales of the problem. Neglecting surface forces, the flow field at the
surface ${\bf \tilde{U}}$, subject to a no-slip condition at the moving
substrate, is given by
\begin{equation}
{\bf \tilde{U}}=-\frac{H^2}{2}\nabla\bar{P}+
H\epsilon\mathrm{Ca}^{-1}\nabla
    \frac{\sigma}{\sigma_0} -\frac{{\bf v_0}}{u_0}\,.
\end{equation}
Here $H=h/h_0$ denotes the nondimensionalized film height, ${\bf v_0}$
stands for the substrate velocity, $\sigma_0$ describes the surface
tension in absence of any surfactant, and $\mathrm{Ca}=\mu u_0 /
\sigma_0$ is the capillary number with dynamic viscosity $\mu$.
Moreover, $\bar{P}$ is a generalized pressure given by
\begin{equation}
\bar{P}=
-\epsilon^3\mathrm{Ca}^{-1}\frac{\sigma}{\sigma_0}\nabla^2 H + \Pi(H)\,.
\end{equation}
One can see that $\bar{P}$ contains, besides the Laplace pressure term
$\sim\nabla^2H$, the disjoining pressure $\Pi(H)$ due to interaction of
substrate and liquid. In the literature, different expressions for the
disjoining pressure have been considered (see
ref.~\cite{ODB_RevModPhys_97} and references therein for a discussion of
possible $\Pi(H)$). Here, we will use the expression
\begin{equation} 
\Pi(H)=\frac{A_n}{H^n}-\frac{A_m}{H^m}\,, \label{e.defpdis}
\end{equation}
with the positive Hamaker constants $A_{n,m}$, where $m>n$, for example
$n=3, m=9$ for Lennard-Jones potentials. The repulsive short-range
interaction prevents a complete dry-off and the substrate is always
covered with a thin precursor film. Gravity could be accounted for by
adding a term $G H$ to $\bar{P}$, but is not considered here since it
plays a minor role in the capillary regime. 

For the height profile of the liquid film, we obtain the standard
evolution equation \cite{ODB_RevModPhys_97}
\begin{equation}
H_{,T}=-\nabla\cdot\left[-\frac{H^3}{3}\nabla\bar{P}+\frac{H^2}{2}\epsilon\mathrm{Ca}^{-1}\nabla\frac{\sigma}{\sigma_0}
-H\frac{{\bf v_0}}{u_0}\right]\,. \label{e.evolh}
\end{equation}
It should be noted, that surfactant thermodynamics affects the
system in two ways. First of all, the diffusive flux ${\bf I}$
is determined by the chemical potential of the surfactant. Second,
the presence of a surfactant alters the surface tension $\sigma$ of
the liquid film, making it dependent on $\gamma$ in a way determined by
ehe equation of state of the surfactant. These two points are discussed
in further details in the following section.
\section{Surfactant thermodynamics}
The lateral pressure $p_\mathrm{lat}$ of a surfactant is defined by
\cite{Adamson}
\begin{equation}
\sigma(\gamma) = \sigma_0 - p_\mathrm{lat}(\gamma)\,.
\end{equation}
Experimentally, $p_\mathrm{lat}(\gamma)$ is usually obtained from
surface tensions measurements using a film balance, where the available
area per surfactant molecule is adjusted with the help of a movable barrier
\cite{KMD_RevModPhys_99,AGS_JPhysique_78,Adamson}. The resulting  isotherms of
material exhibiting the first-order LE/LC transition display a behaviour
reminiscent of a three-dimensional van der Waals gas. It is therefore
reasonable to model the surfactant thermodynamics close to the main
transition by a free energy suitable for a two-dimensional analogue of a
van der Waals gas.
Since the surfactant density varies along the surface, we apply a
Cahn-Hilliard-type free-energy functional \cite{CH_JChemPhys_58},
allowing for a non-uniform free-energy density:
\begin{equation}
\mathcal{F}[\gamma] = \int\mathrm{d}^{D-1}x\left\{
\frac{\kappa}{2}\left(\nabla\gamma\right)^2 +
f_\mathrm{hom}(\gamma) \right\}\,,
\end{equation}
where $\kappa>0$ is constant and $f_\mathrm{hom}(\gamma)$ denotes
the free-energy density of a homogeneous system with surfactant density
$\gamma$.
Assuming the system to be in local thermodynamic equilibrium
the corresponding lateral pressure is given by
\cite{Ev_AdvPhys_79}
\begin{equation}
\label{e.plat}
p_\mathrm{lat}(\gamma)=-f(\gamma) + \gamma\mu^\mathrm{(chem)}(\gamma)\,,
\end{equation}
where the chemical potential $\mu^\mathrm{(chem)}$ is obtained from
$\mathcal{F}$ by functional derivation:
\begin{equation}
\mu^\mathrm{(chem)} = \frac{\delta\mathcal{F}}{\delta\gamma} =
-\kappa\nabla^2\gamma+\frac{\partial f_\mathrm{hom}}{\partial
  \gamma}\,.
\end{equation}
So far, there have been no limitations on the choice of
$f_\mathrm{hom}$. In the spirit of Landau's theory of first-order phase
transitions \cite{LL5} we will now restrict ourselves to free-energy
densities that can be approximated sufficiently well by a fourth-order
polynomial around the critical density $\gamma_{cr}$ of the main
transition. Defining $\tilde{\gamma}=\gamma-\gamma_\mathrm{cr}$ we
obtain
\begin{equation}
f_\mathrm{hom}(\tilde{\gamma})=f_0+f_1\tilde{\gamma}+f_2\tilde{\gamma}^2+f_3\tilde{\gamma}^3+f_4\tilde{\gamma}^4\,.
\end{equation}
Realistic values for the parameters $f_i$ can be estimated by fitting
eq.~\eqref{e.plat} to experimentally obtained isotherms. 
Notice, that $p_\mathrm{lat}$ has to be matched to the measured pressure
within the coexistence region with the help of a Maxwell construction.

The diffusive current ${\bf I}$ in eq.~\eqref{e.evolgamma} is
proportional to the gradient of $\mu^\mathrm{(chem)}$ with proportionality
constant $\alpha$ \cite{LL6}. Hence, in nondimensionalized form, the
lateral pressure and the diffusive current can be written as
\begin{align}
P_\mathrm{lat}= & -\epsilon^2 K
\left[\frac{1}{2}\left(\nabla\Gamma\right)^2 +
\left(\Gamma_\mathrm{cr}+\tilde{\Gamma} \right)\nabla^2\Gamma \right]
\notag \\
& -F_\mathrm{hom}+\left(\Gamma_\mathrm{cr}+\tilde{\Gamma}\right)\frac{\partial F_\mathrm{hom}}{\partial
  \tilde{\Gamma}}\,, \label{e.platdl}
\end{align}
\begin{align}
{\bf I}= & -\epsilon A \frac{\gamma_0}{\sigma_0} \nabla\mu^\mathrm{(chem)} \notag \\ 
= & -\epsilon A\left[-\epsilon^2
K\nabla^3\Gamma+\frac{\partial^2
F_\mathrm{hom}}{\partial \tilde{\Gamma}^2}\nabla\Gamma\right]\,.
\label{e.diffcurrdl}
\end{align}
Here, the dimensionless numbers $A$ and $K$ are defined by
\begin{equation}
A = \frac{\alpha \sigma_0}{h_0 \gamma_0^2 u_0}\,, \qquad K =
\frac{\kappa\gamma_0^2}{h_0^2\sigma_0}\,,
\end{equation}
and the nondimensionalized free-energy density of the homogeneous system
is given by
\begin{equation}
F_\mathrm{hom}=\sum_{n=0}^4 F_n \tilde{\Gamma}^n\,, \quad \textrm{ where } \quad F_n =
\frac{f_n}{\sigma_0}\gamma_0^n \,.
\end{equation}
Inserting eqs.~\eqref{e.platdl} and \eqref{e.diffcurrdl} into the
evolution equations \eqref{e.evolgamma} and \eqref{e.evolh}, we 
obtain the complete set of governing equations:
\begin{align}
H_{,T}= & -\nabla\cdot\left[ \frac{H^3}{3}\nabla \left\{ \epsilon^3\mathrm{Ca}^{-1}\left(
    1-P_\mathrm{lat}(\Gamma)\right)\nabla^2H -\Pi(H) \right\} \right. \notag \\
& \left. -\epsilon\frac{H^2}{2}\mathrm{Ca}^{-1}\nabla
P_\mathrm{lat}(\Gamma) -
H\frac{{\bf v_0}}{u_0} \right] \,, \label{e.Hgov} \\
\Gamma_{,T}= & -\nabla\cdot\left( \Gamma \left[ \frac{H^2}{2}\nabla\left\{
    \epsilon^3\mathrm{Ca}^{-1}\left(
    1-P_\mathrm{lat}(\Gamma)\right) \nabla^2H  \notag \right. \right. \right. \\
& \left. \left. -\Pi(H) \right\} - \epsilon H \mathrm{Ca}^{-1}\nabla P_\mathrm{lat}(\Gamma) - \frac{{\bf v_0}}{u_0}\right] \notag \\
& \left. + \epsilon A \left[\epsilon^2K\nabla^3\Gamma -
\frac{\partial^2F_\mathrm{hom}}{\partial\tilde{\Gamma}^2}\nabla\Gamma
\right] \right) \,. \label{e.Ggov}
\end{align}
In the next section, we will investigate the linear stability of
stationary homogeneous solutions of these equations.
\section{Linear stability analysis}
In the following we concentrate on the case of a substrate at rest,
${\bf v_0}=0$. For the sake of simplicity, we first consider only
one-dimensional fields $H(X,T),\,\Gamma(X,T)$.  Homogeneous film heights
and surfactant densities $H=\hat{H}=\mathrm{const.},
\Gamma=\hat{\Gamma}=\mathrm{const.}$ are always stationary solutions
of the equations. Expanding eqs.~\eqref{e.Hgov} and \eqref{e.Ggov}
like $H(X,T) = \hat{H}+\eta(X,T)$ and
$\Gamma(X,T)=\hat{\Gamma}+\zeta(X,T)$ yields the linearised set of
equations
\begin{equation}
\partial_T\left(\begin{array}{c} \eta \\ \zeta \end{array} \right) =
\mathcal{A}\left(\begin{array}{c} \eta \\ \zeta \end{array}\right)\,,
\end{equation}
with the linear operator 
\begin{equation}
\label{s.long}
\mathcal{A} = \left( \begin{array}{cc}
            \displaystyle \frac{\hat{H}^3}{3}\left(
            \frac{\partial\Pi}{\partial
            H}\bigg|_{\hat{H}}\partial_X^2-\epsilon^3\mathrm{Ca}^{-1}\hat{\Sigma}
\partial_X^4\right)  &
              \displaystyle \epsilon
              \mathrm{Ca}^{-1}\frac{\hat{H}^2}{2}\hat{\Gamma}\left(
                \frac{\partial^2F_\mathrm{hom}}{\partial \tilde{\Gamma}^2}\bigg|_{\Delta\Gamma}\partial_X^2-\epsilon^2K\partial_X^4
                \right) \\
            \displaystyle
            \hat{\Gamma}\frac{\hat{H}^2}{2}\left(\frac{\partial\Pi}{\partial H}\bigg|_{\hat{H}}\partial_X^2
              - \epsilon^3 \mathrm{Ca}^{-1}\hat{\Sigma}\partial_X^4 \right)
                           &
             \displaystyle \epsilon\left(\hat{\Gamma}^2\hat{H}\mathrm{Ca}^{-1}+A\right)\left(
                 \frac{\partial^2F_\mathrm{hom}}{\partial\tilde{\Gamma}^2}\bigg|_{\Delta\Gamma}\partial_X^2-\epsilon^2
                 K \partial_X^4\right) \end{array} \right) 
\end{equation}
where $\hat{\Sigma}=\sigma(\gamma_0\hat{\Gamma})/\sigma_0$ and
$\Delta \Gamma=\hat{\Gamma}-\Gamma_\mathrm{cr}$. Using the
ansatz $\eta\sim\exp(\lambda T + \mathrm{i}kX)$,
       $\zeta\sim\exp(\lambda T + \mathrm{i}kX)$, trace
$\tau$ and determinant $\Delta$ of the resulting matrix
${\bf \mathrm{A}}={\bf \mathrm{A}}(k)$ can be obtained as polynomials of wavenumber
$k$. The growth rates $\lambda$, which are the eigenvalues of
${\bf \mathrm{A}}(k)$, can be calculated by use of
\begin{equation}
\lambda_{\pm} = \frac{\tau \pm
\sqrt{\tau^2-4\Delta}}{2}\,.
\end{equation}
The system is linearly stable if and only if the conditions
$\tau(k)<0$ and $\Delta(k)>0$ are simultaneously fulfilled
for all wavenumbers $k$. Assuming the parameters $\epsilon,
\mathrm{Ca}^{-1}, A, K$ to be positive and taking into account that
$\hat{\Sigma}$ is a surface tension and hence positive as well, the
stability condition can be shown to be equivalent to: 
\begin{equation}
\left\{ \begin{array}{cc} \displaystyle \frac{\partial \Pi}{\partial
  H}\bigg|_{\hat{H}} & > 0\,, \\[10pt] 
    \displaystyle \frac{\partial^2 F_\mathrm{hom}}{\partial
      \tilde{\Gamma}^2}\bigg|_{\Delta\Gamma} & >0\,. \end{array} \right.
      \label{e.stabilitycond}
\end{equation}
In the following analysis, we will use $P_1:=\partial \Pi(\hat{H})
/ \partial H$ and $P_2:=\partial^2F_\mathrm{hom}(\Delta\Gamma)/\partial \Gamma^2$ as control parameters.
Besides the significance of the sign of $P_1$, which is well known from
investigations of spinodal dewetting, there is a similar dependence on
the sign of $P_2$. This is due to the fact, that a homogeneous
distribution of surfactant in the spinodal region, where
$P_2<0$, becomes unstable to spinodal decomposition. 
Although the stability borders are exactly as would be
expected from the isolated subsystems $H$ and $\Gamma$, 
film and surfactant do \emph{not} decouple linearly as will be discussed
in the next section.

An elementary calculation reveals that whenever
condition~\eqref{e.stabilitycond} is violated, there is a band of
unstable modes $0<k<k_\mathrm{c}$, where growth rate
$\mathrm{Re}(\lambda_+)$ is positive. If both, $P_1$ and $P_2$ are
negative, there will also be a band of wavenumbers with positive
$\mathrm{Re}(\lambda_-)$ reaching from $k=0$ to a maximal
wavenumber smaller than $k_\mathrm{c}$. 

In principle, the wavenumber $k_\mathrm{max}$ corresponding to maximal
growth rate $\mathrm{Re}(\lambda_+(k_\mathrm{max}))$ can be calculated
from eq.~\eqref{s.long}, but for the general case the
result cannot be stated in a concise manner. 
However, the upper bound of the band of unstable modes, $k_\mathrm{c}$,
can be calculated analytically and the result depends on the signs of
$P_1$ and $P_2$. Defining
\begin{equation}
k_1=\sqrt{\frac{-P_1}{\epsilon^3\mathrm{Ca}^{-1}\hat{\Sigma}}}, \qquad
k_2=\sqrt{\frac{-P_2}{\epsilon^2K}}, \label{e.konektwo}
\end{equation}  
we obtain
\begin{equation}
k_\mathrm{c} = \left\{ \begin{array}{rcr}
k_1 & \textrm{ for } & P_1<0, P_2>0\,, \\
k_2 & \textrm{ for } & P_1>0,P_2<0\,, \\
\max\left\{k_1, k_2 \right\} & \textrm{ for } & P_1<0,P_2<0 \,.
\end{array}
\right. \label{e.kc}
\end{equation}
This means, that in its lower-left quadrant, the $P_1$-$P_2$ plane is
divided by the line $k_1=k_2$, or equivalently
$P_2=(K/\epsilon\mathrm{Ca}^{-1}\hat{\Sigma})P_1$, into one region where
$k_c=k_1$ and another one, where $k_c=k_2$ (see
fig.~\ref{fig.poneptwo}).

Since operator $\mathcal{A}$ contains only even
powers of $\nabla$, it is clear, that by writing $k:=|{\bf k}|$, the
same results for $\Delta(k)$, $\tau(k)$ and $k_c$ are obtained in the
two-dimensional case, using the ansatz
$\eta\sim\exp(\lambda T + i{\bf k}\cdot{\bf X}),\zeta\sim\exp(\lambda
    T + i{\bf k}\cdot{\bf X})$.
\begin{figure}
\begin{center}
\includegraphics{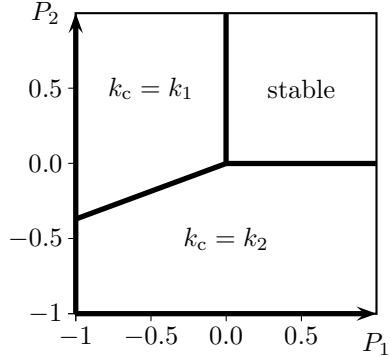}
\caption{Stability diagram of stationary homogeneous solutions in the
  $P_1$-$P_2$ plane. If $P_1$ and $P_2$ are positive, the solution is
    stable. Otherwise, there is a band of unstable modes
    \mbox{$0<k<k_\mathrm{c}$}.}
\label{fig.poneptwo}
\end{center}
\end{figure}
\section{Numerical analysis}
\begin{figure}
\begin{center}
\includegraphics{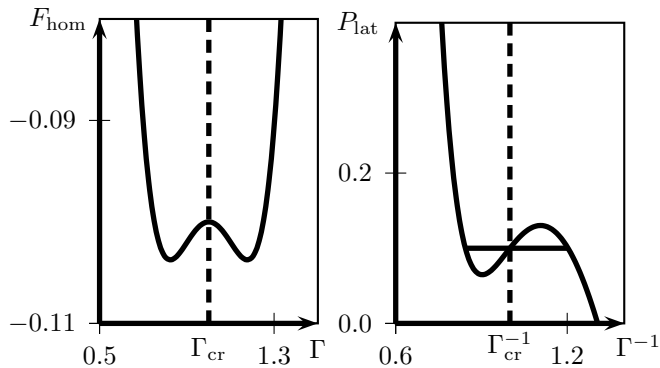}
\caption{The free-energy density (left) used in the numerical example
  and the corresponding pressure-area diagram (right). The solid
    horizontal line results from Maxwell construction.}
\label{fig.potplat}
\end{center}
\end{figure}
\begin{figure}
\begin{center}
\includegraphics{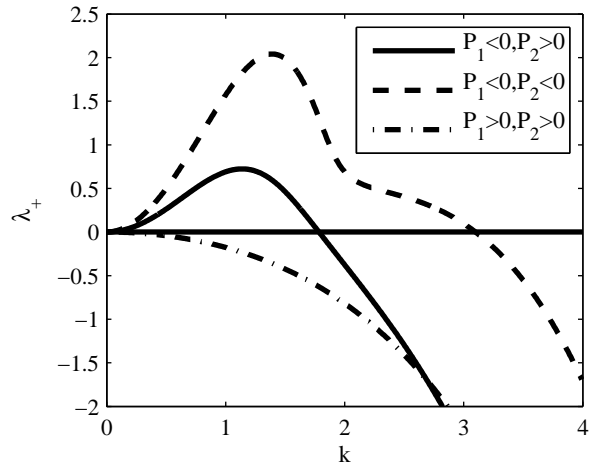}
\caption{Eigenvalue spectra of stability matrix ${\bf \mathrm{A}}$, obtained for
  the parameter set used in our numerical example.}
\label{fig.linstabi}
\end{center}
\end{figure}
We have numerically simulated the nonlinear set of equations
\eqref{e.Hgov} and \eqref{e.Ggov} on periodic domains in one and two
dimensions using a pseudospectral method of lines code \cite{Boyd}.
Time integration was performed by an embedded 4(5) Runge-Kutta scheme,
using the Cash-Karp parameter set \cite{NRC}, while the r.h.s. of
the evolution equations was calculated using 256 Fourier modes in 1D or
$64\times64$ modes in 2D. 

In the simulations we used a simple symmetric double well potential 
$F_\mathrm{hom}$ employing the parameters ${\Gamma_\mathrm{cr}=1},
{F_1=F_3=0}, {F_0=-0.1}, {F_2=-0.24}, {F_4=3.85}$.
Figure~\ref{fig.potplat} shows the free-energy density $F_\mathrm{hom}$,
as is obtained for our choice of parameters, as well as the
resulting pressure-area diagram calculated according to
eq.~\eqref{e.platdl} for homogeneous values of $\Gamma$. The two minima
of $F_\mathrm{hom}$ correspond to two thermodynamically stable phases of
different surfactant density. Within this simple model, the phases of
higher and lower density can be identified with the liquid-condensed (LC)
phase and the liquid-expanded (LE) phase, respectively.

For the disjoining pressure \eqref{e.defpdis} we use the parameters
${n=3},{m=9}$ with ${A_3=3}$ and ${A_9=1}$ and set the remaining
dimensionless numbers to
$\mathrm{Ca}^{-1}=1,A=0.05,K=0.05,{\bf v_0}=0$.
\begin{figure}
\begin{center}
\includegraphics[width=.49\textwidth]{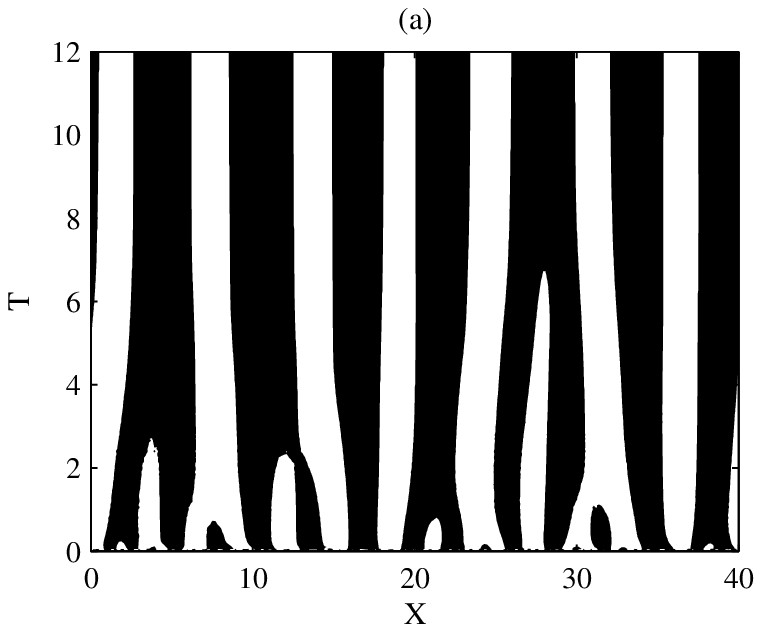}
\includegraphics[width=.49\textwidth]{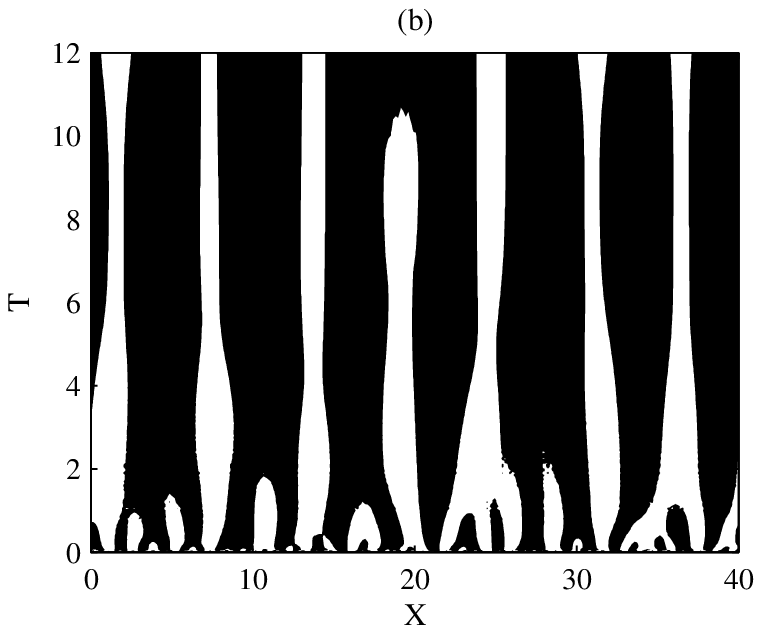}
\caption{Space-time plots of the 1D liquid film $H(X,T)$ (a) and the
  surfactant density $\Gamma(X,T)$ (b), displaying regions of
  positive and negative curvature in black and white, respectively.}
\label{fig.spacetime}
\end{center}
\end{figure}

Our goal is to investigate how the surfactant affects the formation of
droplets. Therefore, as initial conditions, small random perturbations
of the stationary homogeneous solution ${\hat{H}=1.2}$ and
${\hat{\Gamma}=0.82345}$ are used. Since $\partial\Pi(1.2)/\partial H<0$
this corresponds to a flat film, which would be unstable to droplet
formation even in absence of any surfactant. The value of
$\hat{\Gamma}$ is chosen as the position of the lower density minimum of
$F_\mathrm{hom}$. Thus, we are simulating a thin film uniformly covered
with surfactant in its lower density phase. 

The parameter values specified above correspond to the case
$P_1<0,P_2>0$ of the linear stability analysis. The wavenumber
dependent growth rate $\mathrm{Re}(\lambda_+(k))$ is displayed in
fig.~\ref{fig.linstabi}. For comparison we also show
$\mathrm{Re}(\lambda_+(k))$ for two different sets of $\hat{H},
\hat{\Gamma}$. The first one where $P_1<0,P_2<0$ corresponds to
the unstable fixed point $\hat{\Gamma}=\Gamma_\mathrm{cr}=1$ of
$F_\mathrm{hom}$ and the same $\hat{H}=1.2$, whereas the other
describes the stable case $P_1>0,P_2>0$ where again
$\hat{\Gamma}=0.82345$ and $\hat{H}=0.9$.
Now we are in position to determine numerically the
maximal growth rates $\lambda_+(k_\mathrm{max})$ and the corresponding
eigenvectors for the three parameter sets mentioned above.
The calculations show that neither component of the eigenvectors is
dominant. This indicates the coupling of the fields $H$ and $\Gamma$ even
within the scope of linear theory.

Now we concentrate again on the initial condition
$\hat{H}=1.2,\hat{\Gamma}=0.82345$. Time evolution of the
nonlinear system in the one-dimensional case can be described
as follows. In the beginning, spinodal dewetting and surfactant spinodal
decomposition lead to rapid formation of liquid droplets and surfactant
domains, the latter consisting of surfactant in a higher-density (LC)
and a lower-density (LE) phase. Then a coarsening process sets in and
drops of liquid coalesce while surfactant domains merge into larger
ones. 
To visualize the coarsening of the liquid film and the surfactant
density, we display the regions of negative curvature, which naturally
indicate the location of drops and domains of high surfactant density,
in a space-time diagram (see fig.~\ref{fig.spacetime}).
Figure~\ref{fig.film1D} shows a snapshot of a later stage of the 1D
simulation, where the few remaining high-density domains are clearly
located on the largest liquid drops. Evidently, there is a
strong correlation of $\Gamma$ and $H$. In the 2D case, morphology is
similar. As can be seen in fig.~\ref{fig.film2D}, high-density domains
are again located on the largest drops of liquid.
To emphasise this correlation the contour line $\Gamma=1$ is
drawn in the plot of the field $H$ (fig.~\ref{fig.film2D} (a)) and
vice versa. Like in the 1D simulation, drops, that are covered with
surfactant in the low-density phase become smaller and smaller
as $T$ increases, while drops covered with the high-density surfactant
phase grow stronger during the coarsening. Obviously the surfactant has
a sustaining effect on the drops, since the system
energetically favours surfaces with lower surface tension. 
\begin{figure}
\begin{center}
\includegraphics{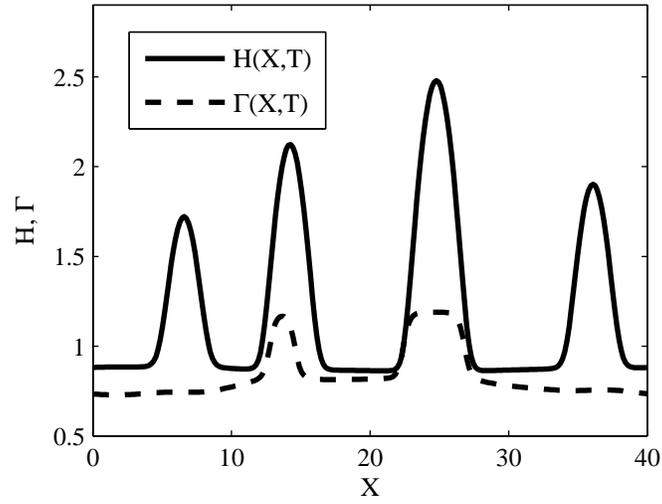}
\caption{One-dimensional film profile (solid) and surfactant density (dashed) at $T=100$.}
\label{fig.film1D}
\end{center}
\end{figure}
\begin{figure}
\begin{center}
\includegraphics[width=.49\textwidth]{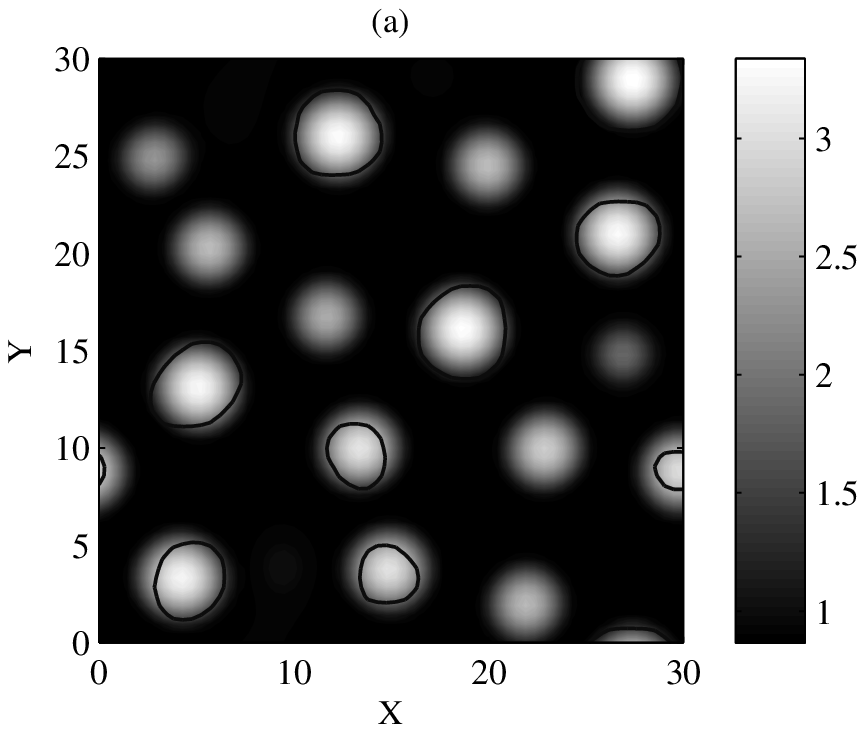}
\includegraphics[width=.49\textwidth]{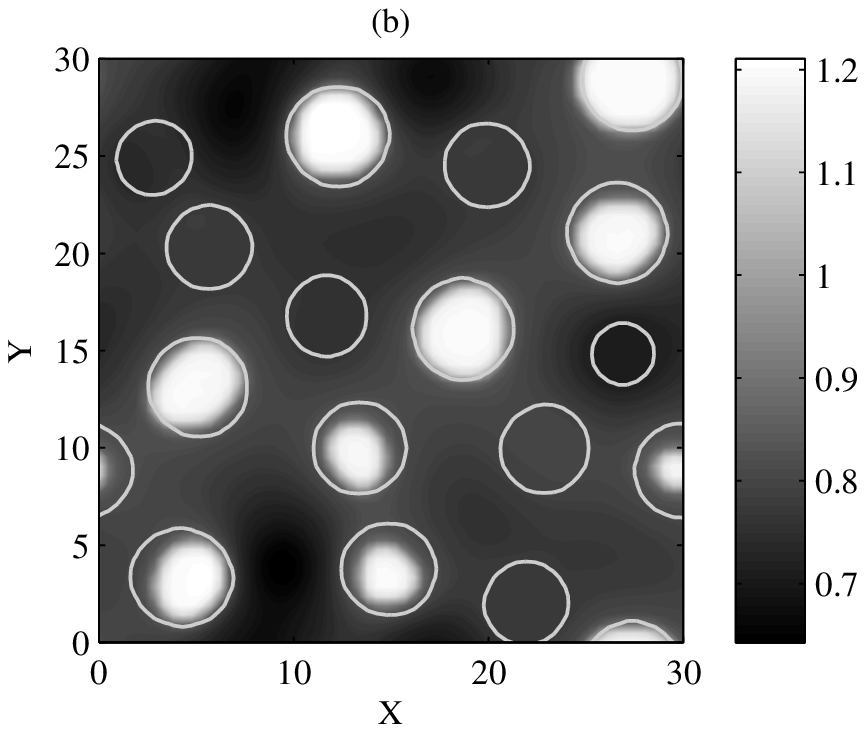}
\caption{Film height (a) and surfactant density (b), each with 
 contour lines ($H,\Gamma=1$) of the other field, at $T=100$.}
\label{fig.film2D}
\end{center}
\end{figure}
\section{Conclusion and outlook}
We have modelled the dynamics of a thin liquid film covered with an
insoluble surfactant in the vicinity of a phase transition. For that
purpose we have incorporated a suitable free-energy functional for the
surfactant into the two governing equations, which were derived within
the lubrication approximation. Linear stability analysis revealed the
interplay of surfactant spinodal decomposition and spinodal dewetting of
the liquid film. Although result \eqref{e.kc}
for $k_c$ seems to imply that the liquid film and the surfactant
are linearly decoupled,
their time evolution is connected from onset. One- and two-dimensional
simulations were presented, showing the decomposition of the surfactant
into domains of material in thermodynamically stable phases. Droplets
and domains show a strong spatial correlation resulting from the
sustaining effect of the surfactant on droplets.  The proposed model
might serve as a starting point for further research on thin film
dynamics with surfactant phase transitions and the related pattern
formation. Also, further investigation is needed, to understand the role
of the phase transition in Langmuir-Blodgett transfer systems.  Thus, it
will be necessary to solve the set of evolution equations subject to
suitable boundary conditions.  
\section{Acknowledgments}
This work was supported by the Deutsche Forschungsgemeinschaft within
special research fund TRR 61. We thank L. F. Chi and M. Hirtz
for helpful discussions.

\end{document}